\documentclass[10pt,twocolumn,conference]{IEEEtran}
\usepackage{setspace}
\usepackage{clrscode}
\usepackage{pict2e}

\usepackage{amsmath}
\usepackage{amssymb}
\usepackage[dvips]{graphicx}
\usepackage{epsfig}
\usepackage{algorithm}
\usepackage{algorithmic}
\usepackage{caption}
\usepackage{subcaption}

\usepackage[ps2pdf,
bookmarks=false,
bookmarksnumbered=false, % true means bookmarks in
bookmarksopen=false, % true means only level 1
colorlinks=false]{}

\newcommand{\E}{\mathbb{E}}

\newcommand{\mH}{\mathcal{H}}
\newcommand{\hH}{\hat{\mathcal{H}}}

\newcommand{\figsize}{0.45}

\newcommand{\nid}{\rm{d}}
\newcommand{\f}{\rm{f}}

\newcommand{\mc}{\rm{c}}
\newcommand{\avg}{\text{avg}}
\newcommand{\pk}{\text{pk}}

\makeatletter

\newcommand{\Rmnum}[1]{\expandafter\@slowromancap\romannumeral #1@}
\makeatother
\newtheorem{Rem}{Remark}

\begin{document}

% paper title
%
\title{Energy-Efficient Power Adaptation for Cognitive Radio Systems under Imperfect Channel Sensing}

\author{\authorblockN{Gozde Ozcan and M. Cenk Gursoy}
\authorblockA{Department of Electrical Engineering and Computer Science\\
Syracuse University, Syracuse, NY 13244\\ Email:
gozcan@syr.edu,
mcgursoy@syr.edu}}
\date{INFOCOM 2014}

\maketitle

\begin{abstract}
In this paper, energy efficient power adaptation is considered in sensing-based spectrum sharing cognitive radio systems in which secondary users first perform channel sensing and then initiate data transmission with two power levels based on the sensing decisions (e.g., idle or busy). It is assumed that spectrum sensing is performed by the cognitive secondary users, albeit with possible errors. In this setting, the optimization problem of maximizing the energy efficiency (EE) subject to peak/average transmission power constraints and average interference constraints is considered. The circuit power is taken into account for total power consumption. By exploiting the quasiconcave property of the EE maximization problem, the original problem is transformed into an equivalent parameterized concave problem and Dinkelbach's method-based iterative power adaptation algorithm is proposed. The impact of sensing performance, peak/average transmit power constraints and average interference constraint on the energy efficiency of cognitive radio systems is analyzed.
\end{abstract}


%\begin{spacing}{1.8}

\section{Introduction}
The significant surge in demand for high data rate wireless applications and the unprecedented growth in the number of wireless users have led to larger amount of bandwidth being required for wireless transmissions and increased the energy consumption levels. On the other hand, high energy prices, limited battery power, increasing greenhouse gas emissions have led to the emerging trend of addressing the optimal and intelligent usage of energy resources. Hence, energy-efficient operation is a major consideration in wireless systems. In addition, bandwidth is generally a scarce resource in wireless communications. Although the available radio-frequency (RF) spectrum has already been allocated/licensed to various applications and services, the allocated spectrum is underutilized most of the time according to the Federal Communication Commission (FCC)'s report \cite{fcc}. This inefficiency in the spectrum usage has led to the consideration of the new communication paradigm of cognitive radio \cite{mitola}, \cite{haykin}. In cognitive radio systems, the unlicensed users (cognitive or secondary users) are able to opportunistically access the frequency bands allocated to the licensed users (primary users) as long as the interference inflicted on the primary users' transmissions is limited. In this regard, cognitive radio enables better and more efficient utilization of the spectrum.

The energy efficiency (EE) of cognitive radio systems has been recently studied. For instance, the authors in \cite{gur} highlight the benefits of cognitive radio systems for green wireless communications. The authors in \cite{pei} design energy efficient optimal sensing strategy and optimal sequential sensing order in multichannel cognitive radio networks. In addition, the sensing time and transmission duration are jointly optimized in \cite{shi}. In the EE analysis of the aforementioned works, secondary users are assumed to transmit only when the channel is sensed as idle. The recent work in \cite{wang} mainly focuses on optimal power allocation to achieve the maximum energy efficiency in OFDM-based cognitive radio networks. Also, energy efficient optimal power allocation in cognitive MIMO broadcast channels is studied in \cite{mao}. In these works, secondary users always share the spectrum with primary users without performing channel sensing.

In order to further increase secondary users' transmission opportunities, unlike above works, in this study we consider the transmission strategy of sensing-based spectrum sharing and assume that the secondary users can coexist with the primary users in the presence of both idle and busy sensing decisions while adapting their transmission power according to the sensing result.
For such a model, we first formulate EE maximization problem subject to peak/average transmit power constraints and average interference constraints in the presence of imperfect sensing results. We explicitly consider circuit power consumption in the total power expenditure. In addition, due to imperfect sensing results, we model the additive disturbance as Gaussian mixture distributed and formulate the achievable rates of the cognitive radio systems accordingly. The EE maximization problem is transformed into an equivalent concave form and Dinkelbach's method-based power allocation algorithm is proposed. We provide numerical results to illustrate the effects of imperfect sensing decisions and transmit/interference power constraints on the energy efficiency.

\section{System Model} \label{sec:system_model}
%We consider sensing-based spectrum sharing cognitive radio system in which a secondary transmitter-receiver pair utilizes the spectrum holes in the licensed bands of the primary users. The term ``spectrum holes" denotes underutilized frequency intervals at a particular time and certain location. In order to detect the spectrum holes,
We assume that the secondary users initially perform channel sensing in the first $\tau$ symbols of the frame duration of $T$ symbols.
%It is assumed that secondary users employ frames of $T$ symbols.
Hence, data transmission is performed in the remaining $T-\tau$ symbols.
\vspace{-.1cm}
\subsection{Channel Sensing}
Spectrum sensing can be formulated as a hypothesis testing problem in which there are two hypotheses based on whether primary users are active or inactive over the channel, denoted by $\mH_{1}$ and $\mH_{0}$, respectively. Many spectrum sensing methods have been studied in the literature (see e.g., \cite{ghasemi}, \cite{axell} and references therein), including matched filter detection, energy detection and cyclostationary feature detection. Each method has its own advantages and disadvantages. However all sensing methods are inevitably subject to errors in the form of false alarms and miss detections due to low signal-to-noise ratio (SNR) of primary users, noise uncertainty, multipath fading and shadowing of wireless channels. Hence, we consider that spectrum sensing is performed with possible errors. The sensing reliability is characterized via only detection and false alarm probabilities. Therefore, any sensing method is applicable in the rest of the analysis. Let $\hH_1$  and $\hH_0$ denote the sensing decisions that the channel is busy (i.e., is occupied by the primary users) and idle, respectively. Hence, by conditioning on the true hypotheses, the detection and false-alarm probabilities are defined, respectively as follows:
\begin{align}
P_{\nid} &= \Pr\{\hH_1 | \mH_1\}, \\
P_{\f} &= \Pr\{\hH_1 | \mH_0\}.
\end{align}
The rest of the conditional probabilities of idle sensing decision given the true hypotheses can be obtained as
\begin{align}
\Pr\{\hH_0 | \mH_1\}=1-P_{\nid}, \\ Pr\{\hH_0 | \mH_0\}=1-P_{\f}.
\end{align}
\subsection{Cognitive Radio Channel Model}
After performing channel sensing, the secondary users initiate data transmission. The channel is considered to be block flat-fading in which the fading coefficients stay the same during a frame duration and vary independently in the following frame. Secondary users are assumed to transmit under both idle and busy sensing decisions. Therefore, as a combination of the true nature of primary user activity and channel sensing decisions, the four possible channel input-output relations between the secondary transmitter-receiver pair can be expressed as follows:
\begin{equation}
\begin{split}
y_i = \begin{cases} hx_{0,i}+ n_i & \text{if } (\mH_{0}, \hH_{0})\\
hx_{1,i}+ n_i  & \text{if } (\mH_{0}, \hH_{1}) \\
hx_{0,i}+ n_i + s_i  & \text{if } (\mH_{1}, \hH_{0}) \\
hx_{1,i}+ n_i + s_i & \text{if } (\mH_{1}, \hH_{1})
\end{cases}
\end{split}
\label{eq:received_signa_BSl}
\normalsize
\end{equation}
where $i= 1, \dots, T-\tau$. Above, $x$  and $y$  are the transmitted and received signals, respectively and $h$ is the channel fading coefficient between the secondary transmitter and the secondary receiver distributed according to a Gaussian distribution with mean zero and variance $\sigma_h^2$. In addition, $n_i$ and $s_i$ denote the additive noise and the primary users' received faded signal. Both $\{n_i\}$ and $\{s_i\}$ are assumed to be independent and identically distributed circularly-symmetric, zero-mean Gaussian sequences with variances $N_0$ and $\sigma_s^2$, respectively. Moreover, the subscripts $0$ and $1$ denote the transmission power levels of the secondary users. More specifically, the average power level is $P_0$ if the channel is detected to be idle while it is $P_1$ if the channel is detected to be busy.

Under these assumptions, the received signal $y$ conditionally given sensing decisions has a Gaussian mixture distribution. In this setting, a closed-form capacity expression is not available. By replacing the conditional distributions with a Gaussian distribution with the same corresponding variance, the following achievable rate (in bits per second) is obtained \cite{ozcan}
\begin{align}
\small
\begin{split}\label{eq:lower_bound_R}
R(P_0,P_1) \!=\!\frac{T\!-\!\tau}{T}\!\sum_{k=0}^1\!\Pr(\hH_k)\E\Bigg\{\!\!\log\!\bigg(\!1\!+\!\frac{P_{k}|h|^2}{N_0\!+\!\Pr(\mH_1|\hH_k)\sigma_s^2}\bigg)\!\Bigg\}
\end{split}
\normalsize
\end{align}
where $\E\{.\}$ denotes expectation operation with respect to the fading coefficient $h$.

The energy efficiency (EE) metric we adopt is the ratio of the achievable rate to the total power consumption (in bits per joule) defined more explicitly as follows:
\begin{align} \label{eq:achievable_EE}
\eta_{\rm{EE}}(P_0,P_1)=\frac{R(P_0,P_1)}{P_{\text{tot}}(P_0,P_1)}=\frac{R(P_0,P_1)}{\Pr\{\hH_0\}P_0+\Pr\{\hH_1\}P_1+P_{\mc}}
\end{align}
Above, the total consumed power consists of average transmission power and circuit power, denoted by $P_{\mc}$. Circuit power represents the average power consumption of the transmitter circuitry (i.e., mixers, filters, and digital-to-analog converters, etc.), which is independent of the transmission power. Also, $\Pr\{\hH_1\}$ and $\Pr\{\hH_0\}$ denote the probabilities of channel being detected as busy and idle, respectively, which can further be expressed as
\begin{equation}
\label{eq:prob_idle_busy}
\begin{split}
\Pr\{\hH_1\} &= \Pr\{\mH_0\}P_{\f}+\Pr\{\mH_1\}P_{\nid},\\
\Pr\{\hH_0\} &= \Pr\{\mH_0\}(1-P_{\f})+\Pr\{\mH_1\}(1-P_{\nid}).
\end{split}
\normalsize
\end{equation}

The achievable EE expression in (\ref{eq:achievable_EE}) can serve as an lower bound since the lower bound on achievable rate $R(P_0,P_1)$ in (\ref{eq:lower_bound_R}) is employed. The usefulness of this EE expression is due to its being an explicit function of the sensing performance.
\begin{figure}[htb]
\centering
\includegraphics[width=\figsize\textwidth]{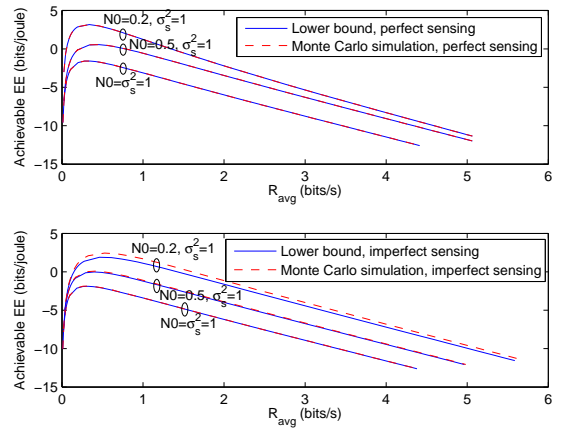}
\caption{Achievable EE $\eta_{\rm{EE}}(P_0,P_1)$ vs. achievable rate $R(P_0,P_1)$.}
\label{fig:EE_SE}
\end{figure}

In Fig. \ref{fig:EE_SE}, we plot the achievable EE expression in (\ref{eq:achievable_EE}) (indicated as the lower bound) and the exact achievable EE as a function of the achievable rate for both perfect sensing (i.e., $P_{\nid} = 1$ and  $P_{\f} = 0$) and imperfect sensing (i.e., $P_{\nid} = 0.8$ and  $P_{\f} = 0.2$). In order to evaluate the exact achievable EE with Gaussian input, we performed Monte Carlo simulations with $2 \times 10^6$ samples. In the case of perfect sensing, the lower bound and simulation result perfectly match as expected since in this case additive disturbance has Gaussian distribution rather than a Gaussian mixture. In the case of imperfect sensing, the lower bound is tight when the difference between noise variance and variance of primary users' is small, e.g., when $N_0=\sigma_s^2=1$ or $N_0=0.5$, $\sigma_s^2=1$. When the difference in the variances is large, e.g., when $N_0=0.2$, $\sigma_s^2=1$, the gap between the lower bound and the exact EE increases. However, it is seen that achievable EE expressions in  (\ref{eq:achievable_EE}) is still a good lower bound. Since circuit power is taken into consideration and we assume $P_c=0.1$, achievable EE vs. achievable rate curve has a bell shape and also is quasiconcave. It is further observed that the maximum EE is attained at nearly the same achievable rate for both lower bound and exact EE expressions.

In the following section, we derive the power adaptation schemes that maximize EE of cognitive radio systems in the presence of sensing errors subject to different combinations of transmit power and interference power constraints.

\section{Optimal Power Adaptation}\label{sec:opt_power}
\subsection{Average Transmit Power Constraint and Average Interference Power Constraint}\label{subsec:average_constraints}
The maximum EE under both average transmit power and interference power constraints can be found by solving the following optimization problem

\small
\begin{align}
\label{eq:EE_avg}
&\max_{
\substack{P_0(g,h) \\ P_1(g,h)}} \eta_{\rm{EE}}(P_0,P_1)\!=\! \frac{R(P_0(g,h),P_1(g,h))}{\E\{\Pr\{\hH_0\}P_0(g,h)\!+\!\Pr\{\hH_1\}P_1(g,h)\}\!+\!P_{\mc}}\\
&\text{subject to} \hspace{0.3cm}\E\{\Pr\{\hH_0\}\,P_0(g,h)  + \Pr\{\hH_1\}\,P_1(g,h)\}  \le P_{\avg} \\
&\hspace{1.6cm}\E\{(1-P_{\nid})\,P_0(g,h)|g|^2 + P_{\nid} \,P_1(g,h)|g|^2\} \le Q_{\avg}\\
&\hspace{1.6cm}P_0(g,h)\geq 0 , P_1(g,h)\geq 0
\end{align}
\normalsize
where $P_{\avg}$ denotes the maximum average transmission power of the secondary transmitter and $Q_{\avg}$ represents the maximum average interference power at the primary receiver. Also, $g$ denotes the channel fading coefficient between the secondary transmitter and the primary receiver and the expectations above are taken with respect to both $g$ and $h$.

The above optimization problem is quasiconcave since the achievable rate $R(P_0,P_1)$ is concave in transmission powers, and the total power consumption $P_{\text{tot}}(P_0,P_1)$ is both affine and positive. Then, the level sets $S_{\alpha}=\{P_0, P_1 : \eta_{\rm{EE}}(P_0,P_1)\geq \alpha\}=\{\alpha P_{\text{tot}}(P_0,P_1)- R(P_0,P_1) \le 0\}$ are convex for any $\alpha \in \mathbb{R}$.
%Since quasiconcave functions have more than one local maximum, local maximum does not always guarantee the global maximum. Therefore, standard convex optimization algorithms cannot be directly used.
We employ an iterative power adaptation algorithm based on Dinkelbach's method \cite{dinkelbach} to solve the quasiconcave EE maximization problem by considering the equivalent parameterized concave problem as follows:

\small
\begin{align}\label{eq:parametrized_EE} \nonumber
& \max_{
\substack{P_0(g,h) \\ P_1(g,h)}}\Big\{R(P_0(g,h),P_1(g,h))-\alpha (\E\{\Pr\{\hH_0\}P_0(g,h)\\&\hspace{4.5cm}+\!\Pr\{\hH_1\}P_1(g,h)\}\!+\!P_{\mc})\Big\}
\end{align}
\normalsize
\begin{align}
\label{eq:Pavg_cons}
&\hspace{-0.2cm}\text{subject to} \hspace{0.3cm}\E\{\Pr\{\hH_0\}\,P_0(g,h)  + \Pr\{\hH_1\}\,P_1(g,h)\}  \le P_{\avg} \\ \label{eq:Qavg_cons}
&\hspace{0.7cm}\E\{(1-P_{\nid})\,P_0(g,h)|g|^2 + P_{\nid} \,P_1(g,h)|g|^2\} \le Q_{\avg}\\
&\hspace{0.7cm}P_0(g,h)\geq 0 , P_1(g,h)\geq 0
\end{align}
\normalsize
where $\alpha$ is a nonnegative parameter. At the optimal value of $\alpha^*$, solving the EE maximization problem in (\ref{eq:EE_avg}) is equivalent to solving the above parametrized concave problem if and only if the following condition is satisfied
\begin{equation}
\begin{split} \label{eq:cond_opt}
F(\alpha^*)=R(P_0(g,h),P_1(g,h))-\alpha^*& (\E\{\Pr\{\hH_0\}P_0(g,h)\\&\hspace{-0.7cm}+\!\Pr\{\hH_1\}P_1(g,h)\}\!+\!P_{\mc})=0
\end{split}
\end{equation}
The detailed proof of the above condition is available in \cite{dinkelbach}. Since the problem in (\ref{eq:parametrized_EE}) is concave, the optimal power values are obtained by forming the Lagrangian function as follows
\begin{equation}
\begin{split}
&\hspace{-0.1cm}L(P_0,P_1,\lambda,\nu,\alpha)=R(P_0(g,h),P_1(g,h))\\&\hspace{-0.15cm}-\alpha (\E\{\Pr\{\hH_0\}P_0(g,h)+\Pr\{\hH_1\}P_1(g,h)\}\!+\!P_{\mc})\\&\hspace{-0.15cm}-\lambda (\E\{\Pr\{\hH_0\}\,P_0(g,h)  + \Pr\{\hH_1\}\,P_1(g,h)\} -P_{\avg} )\\&\hspace{-0.15cm}-\nu (\E\{(1-P_{\nid})\,P_0(g,h)|g|^2 + P_{\nid} \,P_1(g,h)|g|^2\} -Q_{\avg})
\end{split}
\end{equation}
\normalsize
where $\lambda$ and $\nu$ are nonnegative Lagrangian multipliers. According to the Karush-Kuhn-Tucker (KKT) conditions, the optimal values of $P_0^*(g,h)$ and $P_1^*(g,h)$ satisfy the following equations

\small
\begin{align}
&\hspace{-0.3cm}\frac{\frac{T-\tau}{T}\Pr\{\hH_0\}|h|^2\log_2e}{N_0\!+\!\Pr(\mH_1|\hH_0)\sigma_s^2\!+\!P_0^*(g,h)|h|^2}\!-\!(\lambda\!+\!\alpha) \Pr\{\hH_0\} \!-\! \nu |g|^2(1\!-\!P_{\nid})\!=\!0 \\
\hspace{-0.6cm}&\frac{\frac{T-\tau}{T}\Pr\{\hH_1\}|h|^2\log_2e}{N_0\!+\!\Pr(\mH_1|\hH_1)\sigma_s^2\!+\!P_1^*(g,h)|h|^2}-(\lambda\!+\!\alpha)  \Pr\{\hH_1\} - \!\nu |g|^2P_{\nid}\!=\!0\\
&\lambda (\E\{\Pr\{\hH_0\}\,P_0^*(g,h)  + \Pr\{\hH_1\}\,P_1^*(g,h)\}  - P_{\avg}) =0\\
& \nu (\E\{(1-P_{\nid})\,P_0^*(g,h)|g|^2 + P_{\nid} \,P_1^*(g,h)|g|^2\} -Q_{\avg})=0 \\
& \lambda \geq 0, \nu \geq 0
\end{align}
\normalsize
Hence, the optimal power values $P_0^*(g,h)$ and $P_1^*(g,h)$ can be found, respectively as in (\ref{eq:opt_P0_avg}) and (\ref{eq:opt_P1_avg}) given at the top of the next page, where $[x]^+$ denotes $\max(x,0)$.
\begin{figure*}
\begin{align} \label{eq:opt_P0_avg}
&P_0^*(g,h)=\Bigg[\frac{\frac{T-\tau}{T}\Pr\{\hH_0\}\log_2e}{\lambda \Pr\{\hH_0\} +\nu |g|^2(1-P_{\nid})+(\lambda+\alpha)\Pr\{\hH_0\}}-\frac{N_0+\Pr(\mH_1|\hH_0)\sigma_s^2}{|h|^2}\Bigg]^+\\\label{eq:opt_P1_avg}
& P_1^*(g,h)=\Bigg[\frac{\frac{T-\tau}{T}\Pr\{\hH_1\}\log_2e}{\lambda \Pr\{\hH_1\} +\nu |g|^2P_{\nid}+(\lambda+\alpha)\Pr\{\hH_1\}}-\frac{N_0+\Pr(\mH_1|\hH_1)\sigma_s^2}{|h|^2}\Bigg]^+
\end{align}
\hrule
\end{figure*}
The Lagrange multipliers $\lambda$ and $\nu$ can be jointly obtained by inserting the optimal power adaptation formulations in (\ref{eq:opt_P0_avg}) and (\ref{eq:opt_P1_avg}) into the constraints given in (\ref{eq:Pavg_cons}) and (\ref{eq:Qavg_cons}). However, solving these equations does not result in closed form expressions for $\lambda$ and $\nu$. Therefore, subgradient method is employed, i.e., $\lambda$ and $\nu$ are updated iteratively according to the subgradient direction until convergence as follows
\begin{align}
&\hspace{-0.1cm}\lambda^{(\!n\!+\!1\!)}\!\!=\!\!\Big[\!\lambda^{(\!n\!)}\!\!-\!t\big( \!P_{\avg}\!-\!\E\!\{\Pr\{\!\hH_0\}\!\,P_0^{(\!n\!)}(\!g,\!h)  \!+\! \Pr\{\!\hH_1\}\!\,P_1^{(\!n\!)}(g,\!h)\}\!\big)\!\Big]^+ \\ \label{eq:nu_update}
&\hspace{-0.1cm}\nu^{(\!n\!+\!1\!)}\!=\!\!\Big[\nu^{(\!n\!)}\!\!-\!t\big( Q_{\avg}\!-\!\E\{\big((1\!-\!P_{\nid})\!\,P_0^{(\!n\!)}\!(g,h)\! +\! P_{\nid} \!\,P_1^{(\!n\!)}(g,h)\big)|g|^2\}\!\big)\Big]^+
\end{align}
where $n$ denotes the iteration index and $t$ denotes the step size. When the step size is chosen to be constant, it was shown that the subgradient method is guaranteed to converge to the optimal value within a small range \cite{boyd}.

For a given value of $\alpha$, the optimal power adaptations in (\ref{eq:opt_P0_avg}) and (\ref{eq:opt_P1_avg}) are found until $F(\alpha) \le \epsilon$ is satisfied. Dinkelbach's method converges to the optimal solution at a superlinear convergence rate. The detailed proof of convergence can be found in \cite{schaible}. In the case of $F(\alpha)=0$ in (\ref{eq:cond_opt}), the solution is optimal. Otherwise, an $\epsilon$-optimal solution is obtained. In the following table, Dinkelbach's method-based iterative power adaptation algorithm for energy efficiency maximization under imperfect sensing is summarized.

\begin{algorithm}
    \caption{Dinkelbach's method-based power adaptation that maximizes the EE of cognitive radio systems under both average transmit power and interference constraints}
    \begin{algorithmic}[1]
      \STATE Initialization: $P_{\nid}=P_{\nid,init}$, $P_{\f}=P_{\f,init}$, $\epsilon > 0$, $t > 0$, $\alpha^{(0)}=\alpha_{\text{init}}$,  $\lambda^{(0)}=\lambda_{\text{init}}$, $\nu^{(0)}=\nu_{\text{init}}$
      \STATE $n \leftarrow 0$
      \REPEAT
       \STATE calculate $P_0^*(g,h)$ and $P_1^*(g,h)$ using (\ref{eq:opt_P0_avg}) and (\ref{eq:opt_P1_avg}), respectively;
	\STATE update $\lambda$ and $\nu$ using subgradient method as follows;
        \STATE $k \leftarrow 0$
      \REPEAT
       \STATE $\lambda^{(k+1)}\!\!\!=\Big[\lambda^{(k)}\!\!-t\big( P_{\avg}\!-\!\E\{\Pr\{\hH_0\}\!\,P_0^{(\!k\!)}(g,\!h)  + \Pr\{\hH_1\}\!\,P_1^{(\!k\!)}(g,\!h)\}\big)\Big]^+$
	\STATE $\nu^{(k+1)}=\Big[\nu^{(k)}-\!t\big( Q_{\avg}-\E\{(1\!-\!P_{\nid})\,P_0^{(\!k\!)}(g,h)|g|^2\! +\! P_{\nid} \!\,P_1^{(\!k\!)}(g,h)|g|^2\}\big)\Big]^+$
\STATE $k \leftarrow k+1$
       \UNTIL{$|\nu^{(k)}\big( Q_{\avg}\!-\!\E\{(1-P_{\nid})\!\,P_0^{(\!k\!)}(g,h)|g|^2\! +\! P_{\nid} \!\,P_1^{(\!k\!)}(g,h)|g|^2\}\big)| \hspace{-0.1cm}\le \epsilon$ and $|\lambda^{(k)}\big( P_{\avg}\!\!\!\!-\!\E\{\Pr\{\hH_0\}\!\,P_0^{(\!k\!)}(g,\!h)  \!+\! \Pr\{\hH_1\}\!\,P_1^{(\!k\!)}(g,\!h)\}\big)| \le \epsilon$}
\STATE $\alpha^{(n+1)}=\frac{R(P_0^*(g,h),P_1^*(g,h))}{\E\{\Pr\{\hH_0\}P_0^*(g,h)\!+\!\Pr\{\hH_1\}P_1^*(g,h)\}\!+\!P_{\mc}}$
\STATE $n \leftarrow n+1$
       \UNTIL{$|F(\alpha^{(n)})| \le \epsilon$}
    \end{algorithmic}
  \end{algorithm}

Note that in the case of $\alpha=0$, EE maximization problem is equivalent to spectral efficiency (SE) maximization.

\begin{Rem}
The power adaptation schemes in (\ref{eq:opt_P0_avg}) and (\ref{eq:opt_P1_avg}) depend on the channel quality between the secondary transmitter and secondary receiver, denoted by $|h|^2$, the interference channel quality between the secondary transmitter and the primary receiver, $|g|^2$, and the sensing performance through detection and false alarm probabilities, $P_{\nid}$ and $P_{\f}$, respectively. When both perfect sensing, i..e., , $P_{\nid}=1$ and $P_{\f}=0$, and SE maximization are considered, i.e., $\alpha$ is set to $0$, the power adaptation schemes become similar to that given in \cite{kang}. However, the secondary users have two power adaptation schemes depending on the presence or absence of the primary users.
\end{Rem}

\subsection{Peak Transmit Power Constraint and Average Interference Power Constraint}
Next, we consider peak transmit power constraint and average interference constraint for EE maximization in cognitive radio systems. In this case, energy-efficient power adaptation can be obtained by solving the following problem:

\small
\begin{align}
%\label{eq:EE_avg}
&\hspace{-0.2cm}\max_{
\substack{P_0(g,h) \\ P_1(g,h)}} \eta_{\rm{EE}}(P_0,P_1)\!=\! \frac{R(P_0(g,h),P_1(g,h))}{\E\{\Pr\{\hH_0\}P_0(g,h)\!+\!\Pr\{\hH_1\}P_1(g,h)\}\!+\!P_{\mc}}\\
&\hspace{-0.2cm}\text{subject to} \hspace{0.2cm} P_0(g,h)  \le P_{\pk,0} \\
&\hspace{1.3cm} P_1(g,h)  \le P_{\pk,1} \\
&\hspace{1.3cm}\E\{(1-P_{\nid})\,P_0(g,h)|g|^2 + P_{\nid} \,P_1(g,h)|g|^2\} \le Q_{\avg}\\
&\hspace{1.3cm}P_0(g,h)\geq 0 , P_1(g,h)\geq 0
\end{align}
\normalsize
where $P_{\pk,0}$ and $P_{\pk,1}$ denote the peak transmit power limits when the channel is detected as idle or busy, respectively.

By transforming the above optimization problem into an equivalent parametrized concave form and following the same steps as in Section \ref{subsec:average_constraints}, the power adaptation schemes are obtained as in (\ref{P0_peak}) and (\ref{P1_peak}), respectively, given at the top of the next page.

\begin{figure*}
\begin{equation} \label{P0_peak}
\begin{split}
P_0^*(g,h) = \begin{cases} 0, &|g|^2 \geq \overbrace{\frac{\Pr\{\hH_0\}}{\nu (1-P_{\nid})} \frac{\frac{T-\tau}{T}\log_2e |h|^2}{N_0+\Pr(\mH_1|\hH_0)\sigma_s^2}-\alpha\Pr\{\hH_0\}}^{g_1}\\
\frac{\frac{T-\tau}{T}\Pr\{\hH_0\}\log_2e}{\nu |g|^2(1-P_{\nid})+\alpha\Pr\{\hH_0\}}-\frac{N_0+\Pr(\mH_1|\hH_0)\sigma_s^2}{|h|^2},  &g_1> |g|^2 > \underbrace{\frac{\Pr\{\hH_0\}}{\nu (1-P_{\nid})} \frac{\frac{T-\tau}{T}\log_2e}{P_{\pk,0}+N_0+\Pr(\mH_1|\hH_0)\sigma_s^2}-\alpha\Pr\{\hH_0\}}_{g_2} \\
P_{\pk,0}, &|g|^2 \le g_2
\end{cases} \\
\end{split}
\normalsize
\end{equation}
\begin{equation}\label{P1_peak}
\hspace{-0.9cm}
\begin{split}
P_1^*(g,h) = \begin{cases} 0, &|g|^2 \geq \overbrace{\frac{\Pr\{\hH_1\}}{\nu P_{\nid}} \frac{\frac{T-\tau}{T}\log_2e |h|^2}{N_0+\Pr(\mH_1|\hH_1)\sigma_s^2}-\alpha\Pr\{\hH_1\}}^{\hat{g}_1}\\
\frac{\frac{T-\tau}{T}\Pr\{\hH_1\}\log_2e}{\nu |g|^2 P_{\nid}+\alpha \Pr\{\hH_1\}}-\frac{N_0+\Pr(\mH_1|\hH_1)\sigma_s^2}{|h|^2},  &\hat{g}_1> |g|^2 > \underbrace{\frac{\Pr\{\hH_1\}}{\nu P_{\nid}} \frac{\frac{T-\tau}{T}\log_2e}{P_{\pk,1}+N_0+\Pr(\mH_1|\hH_1)\sigma_s^2}-\alpha\Pr\{\hH_1\}}_{\hat{g}_2} \\
P_{\pk,1}, &|g|^2 \le \hat{g}_2
\end{cases}
\end{split}
\normalsize
\end{equation}
\hrule
\end{figure*}
\begin{Rem}
The power adaptation schemes in (\ref{P0_peak}) and (\ref{P1_peak}) has the same structure as those in \cite{kang} in the case of perfect sensing and $\alpha=0$.
\end{Rem}
Algorithm 1 can be modified to maximize the EE subject to peak power constraint and average interference constraint in such a way that $P_0^*(g,h)$ and $P_1^*(g,h)$ are calculated using (\ref{P0_peak}) and (\ref{P1_peak}), respectively, and only the Lagrange multiplier $\nu$ is updated according to (\ref{eq:nu_update}).

\section{Numerical Results}\label{sec:num_results}
In this section, we present numerical results to illustrate the performance of the proposed EE-maximizing power adaptation methods. Unless mentioned explicitly, it is assumed that noise variance is $N_0=0.2$, the variance of primary user signal is $\sigma_s^2=1$. Also, the prior probabilities $\Pr\{\mH_0\} = 0.4 $ and $\Pr\{\mH_1\} = 0.6$. The frame duration $T$ and sensing duration $\tau$ are set to $100$ and $10$, respectively. The circuit power is $P_c = 0.1$. The step sizes  $\lambda$ and $\nu$ are set to $0.1$ and tolerance $\epsilon$ is chosen as $0.0001$.

\begin{figure}[htb]
\centering
\includegraphics[width=\figsize\textwidth]{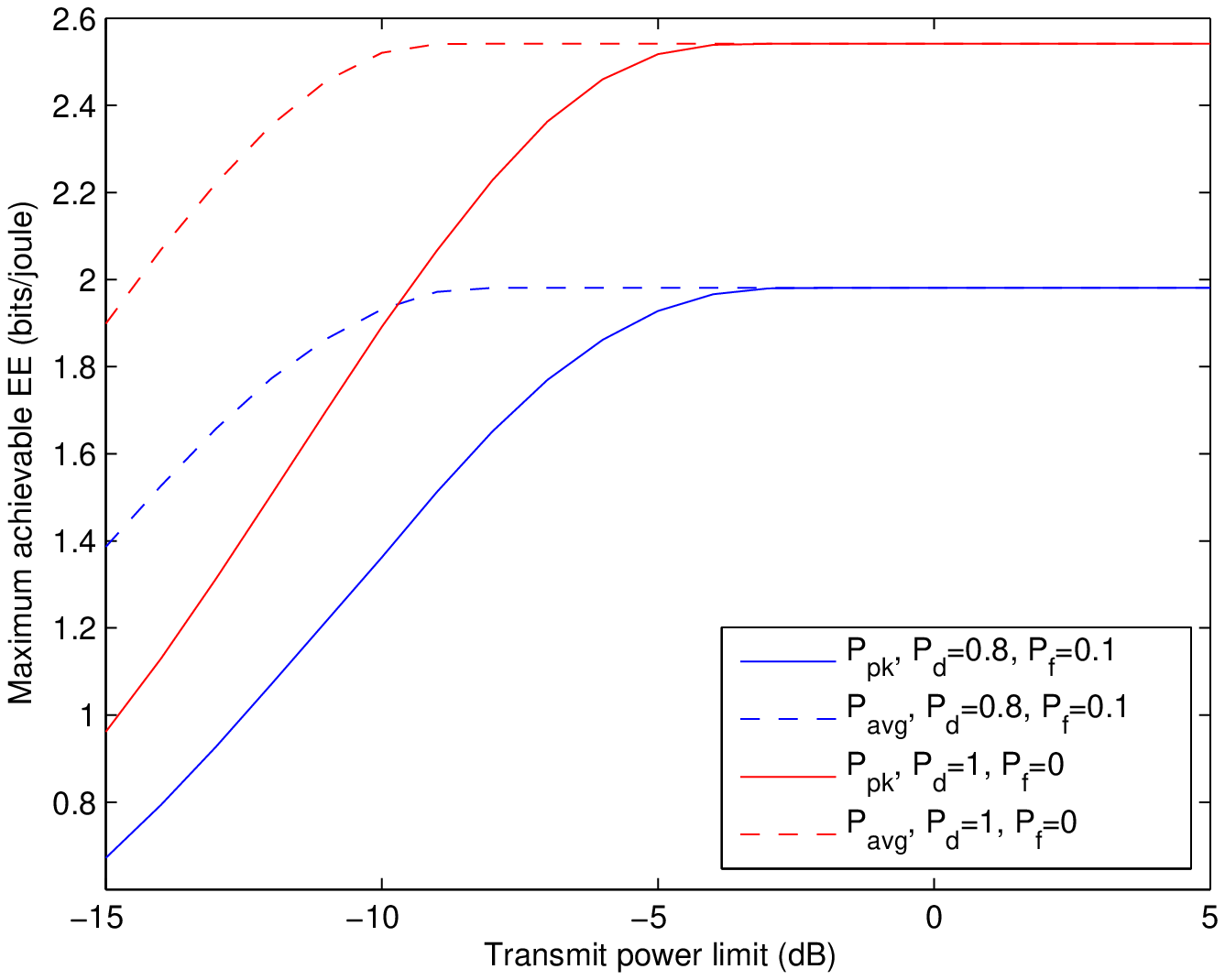}
\caption{Achievable EE $\eta_{\rm{EE}}(P_0,P_1)$ vs. peak/average transmit power constraints.}
\label{fig:EE_transmitpower_limit}
\end{figure}

\begin{figure*}[htb]
\centering
\begin{subfigure}[b]{0.32\textwidth}
\centering
\includegraphics[width=\textwidth]{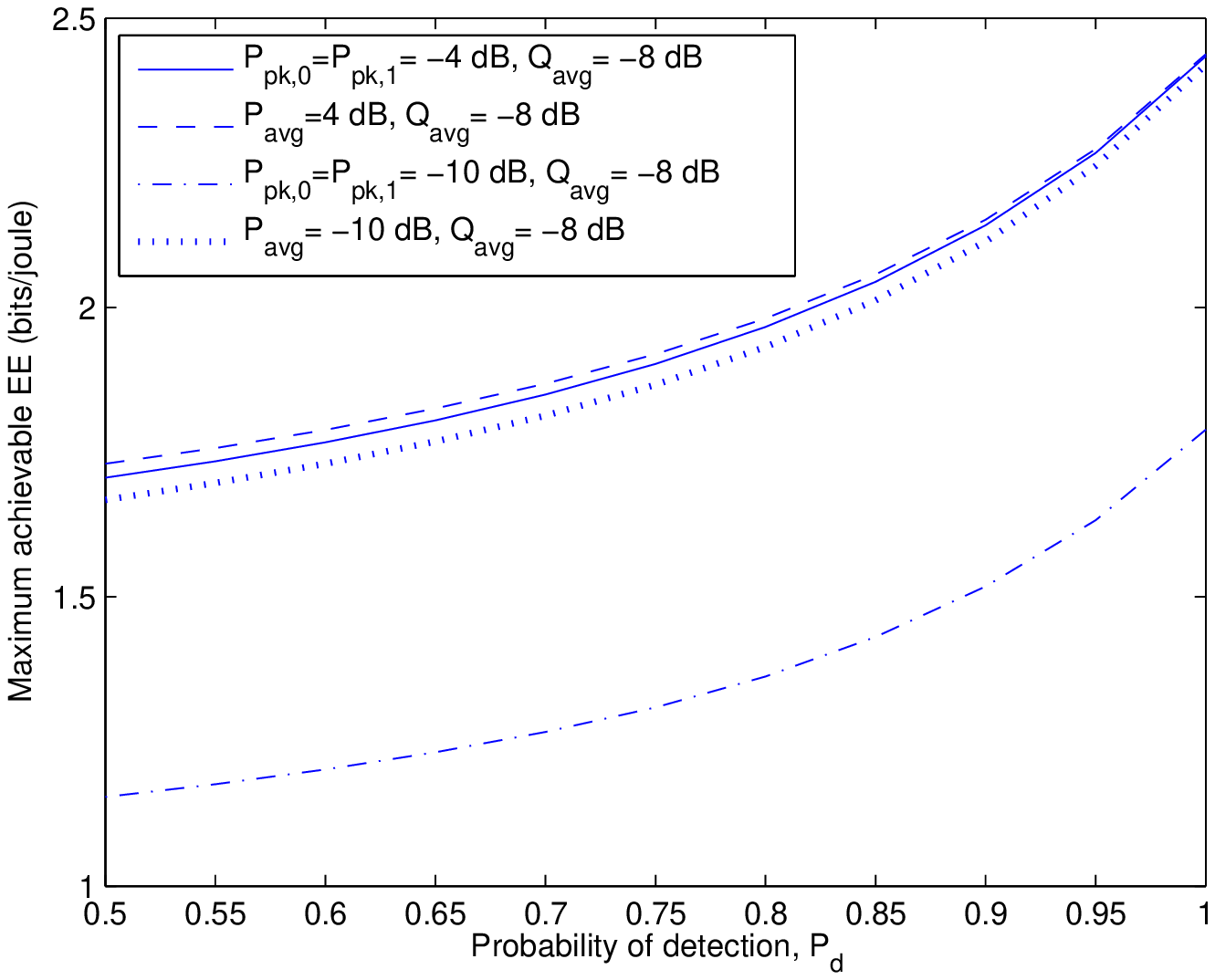}
\caption{$\eta_{\rm{EE}}(P_0,P_1)$ vs. $P_{\nid}$}
\end{subfigure}
\begin{subfigure}[b]{0.32\textwidth}
\centering
\includegraphics[width=\textwidth]{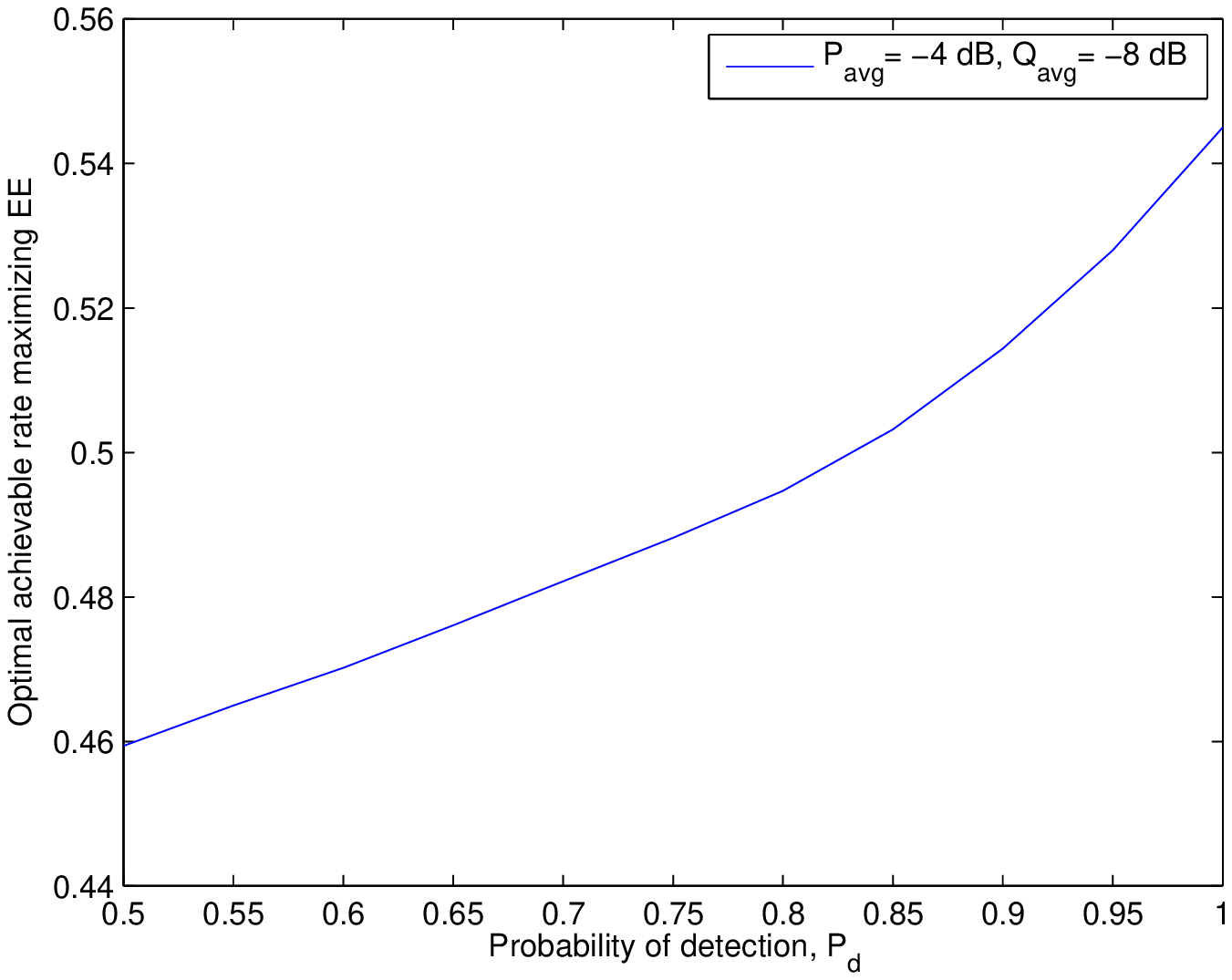}
\caption{$R(P_0,P_1)$ vs. $P_{\nid}$}
\end{subfigure}
\begin{subfigure}[b]{0.32\textwidth}
\centering
\includegraphics[width=\textwidth]{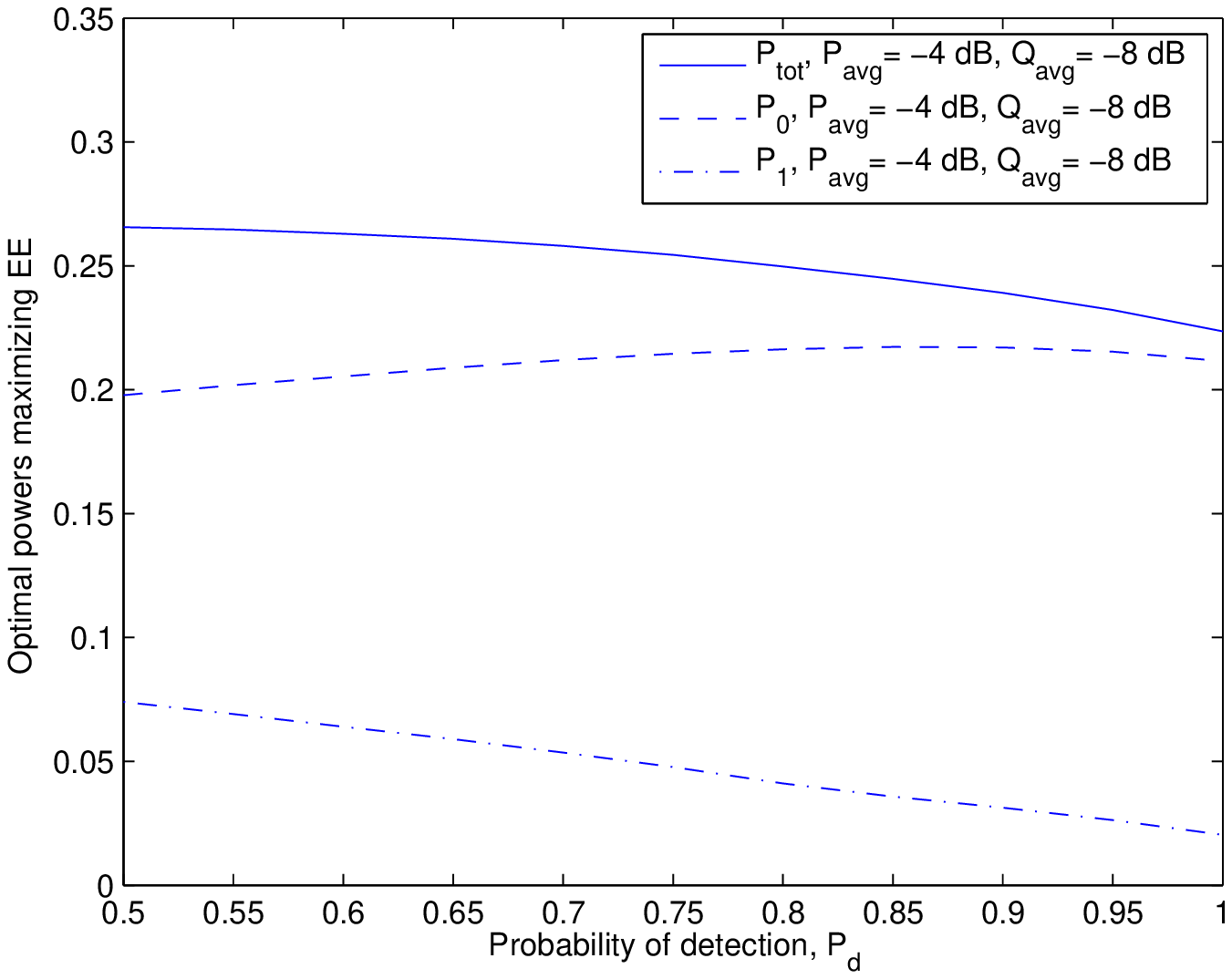}
\caption{$P_{\text{tot}}$, $P_0$ and $P_1$ vs. $P_{\nid}$}
\end{subfigure}
\caption{\small{(a) Maximum achievable EE, $\eta_{\rm{EE}}(P_0,P_1)$ vs. probability of detection, $P_{\nid}$; (b) optimal achievable rate maximizing EE, $R(P_0,P_1)$ vs. $P_{\nid}$; (c) optimal total transmission power, $P_{\text{tot}}$ and $P_0$, $P_1$ vs. $P_{\nid}$.}}\label{fig:EE_Pd}
\end{figure*}

In Fig. \ref{fig:EE_transmitpower_limit}, we display achievable maximum EE as a function of the constraints on peak/average transmit power for perfect sensing (i.e., $P_{\nid}=1$ and $P_{\f}=0$) and imperfect sensing with $P_{\nid}=0.8$ and $P_{\f}=0.1$. $Q_{\avg}$ is set to $-1$ dB. It is seen that higher energy efficiency is achieved with perfect sensing compared to that attained with imperfect sensing. In the case of perfect sensing, the probabilities $\Pr(\mH_1|\hH_0)$ and $\Pr(\mH_0|\hH_1)$ are zero. Therefore, the secondary users in idle-sensed channels do not experience additive disturbance from the primary users, which results in higher achievable rates, hence higher achievable EE compared to that in the imperfect-sensing case.  It is also observed that maximum achievable EE initially increases as the peak/average transmit power constraints relax. However, when the peak/average transmit power constraints become sufficiently looser compared to $Q_{\avg}$, the maximum achievable EE becomes fixed since the transmission power is now determined by the average interference constraint, $Q_{\avg}$ rather than the peak/average transmit power constraints. Moreover, higher achievable EE is achieved under the average transmit power constraint since the power allocation under the average transmit power constraint is more flexible than that under the peak transmit power constraint.

\begin{figure*}[htb]
\centering
\begin{subfigure}[b]{0.32\textwidth}
\centering
\includegraphics[width=\textwidth]{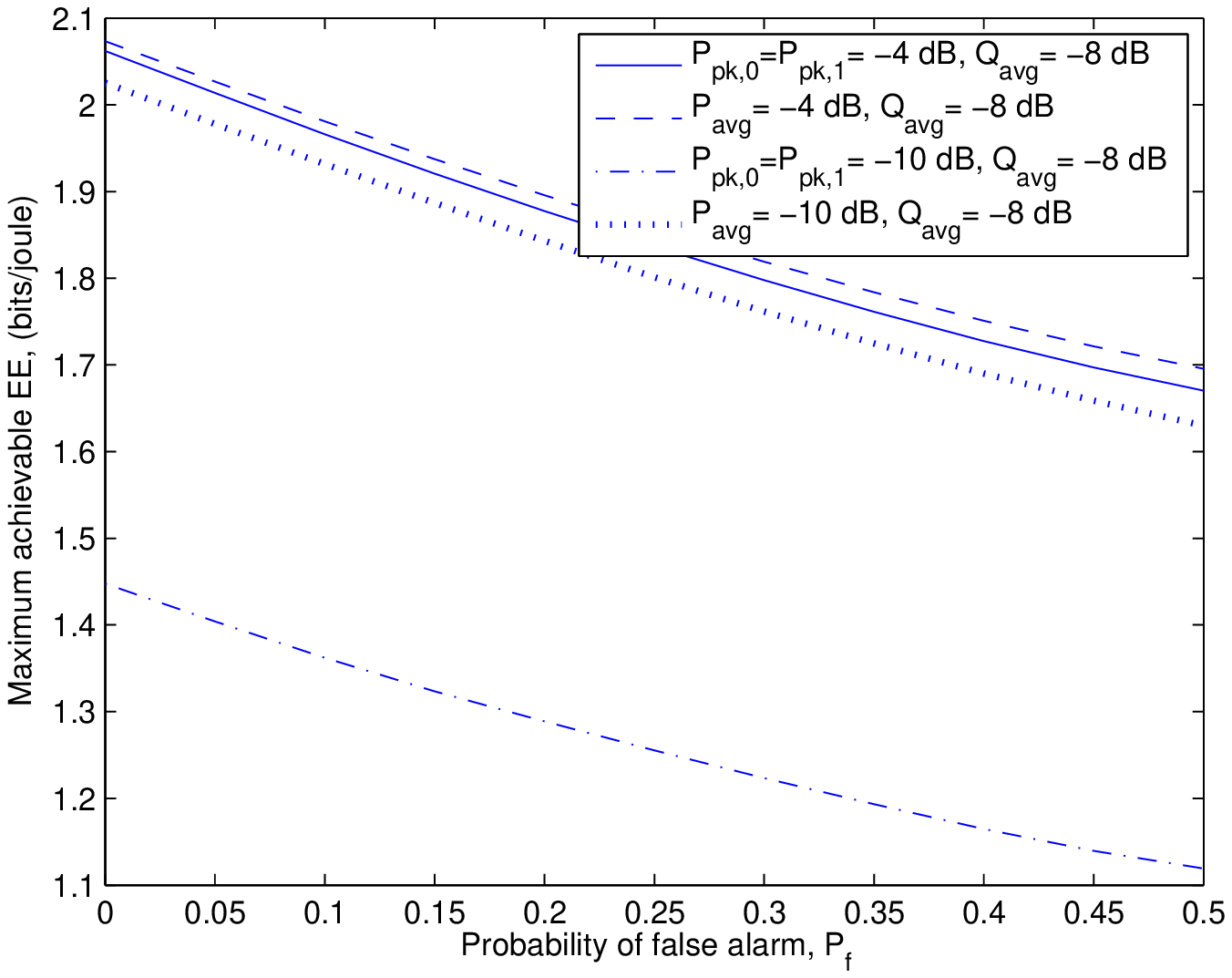}
\caption{$\eta_{\rm{EE}}(P_0,P_1)$ vs. $P_{\f}$}
\end{subfigure}
\begin{subfigure}[b]{0.32\textwidth}
\centering
\includegraphics[width=\textwidth]{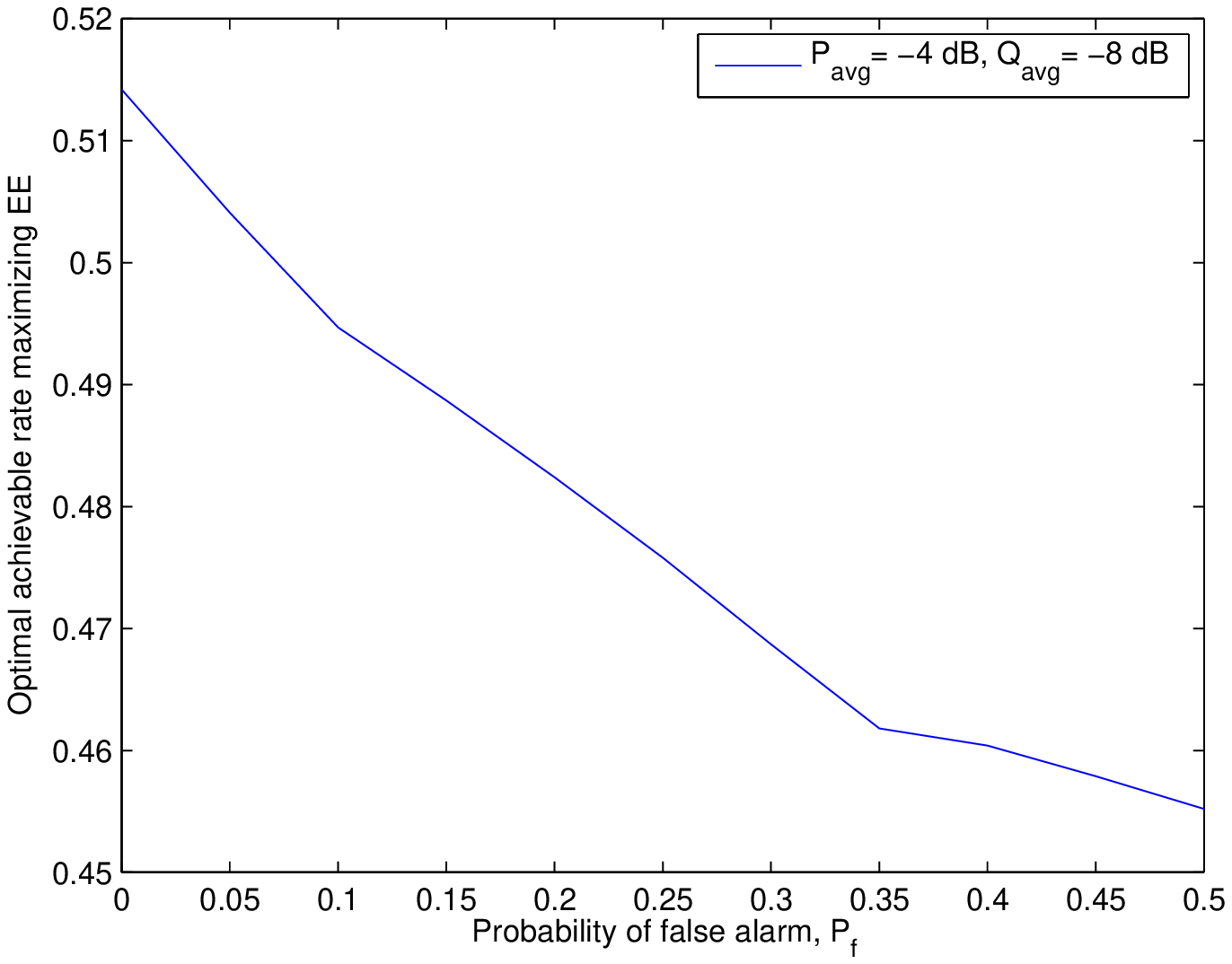}
\caption{$R(P_0,P_1)$ vs. $P_{\f}$}
\end{subfigure}
\begin{subfigure}[b]{0.32\textwidth}
\centering
\includegraphics[width=\textwidth]{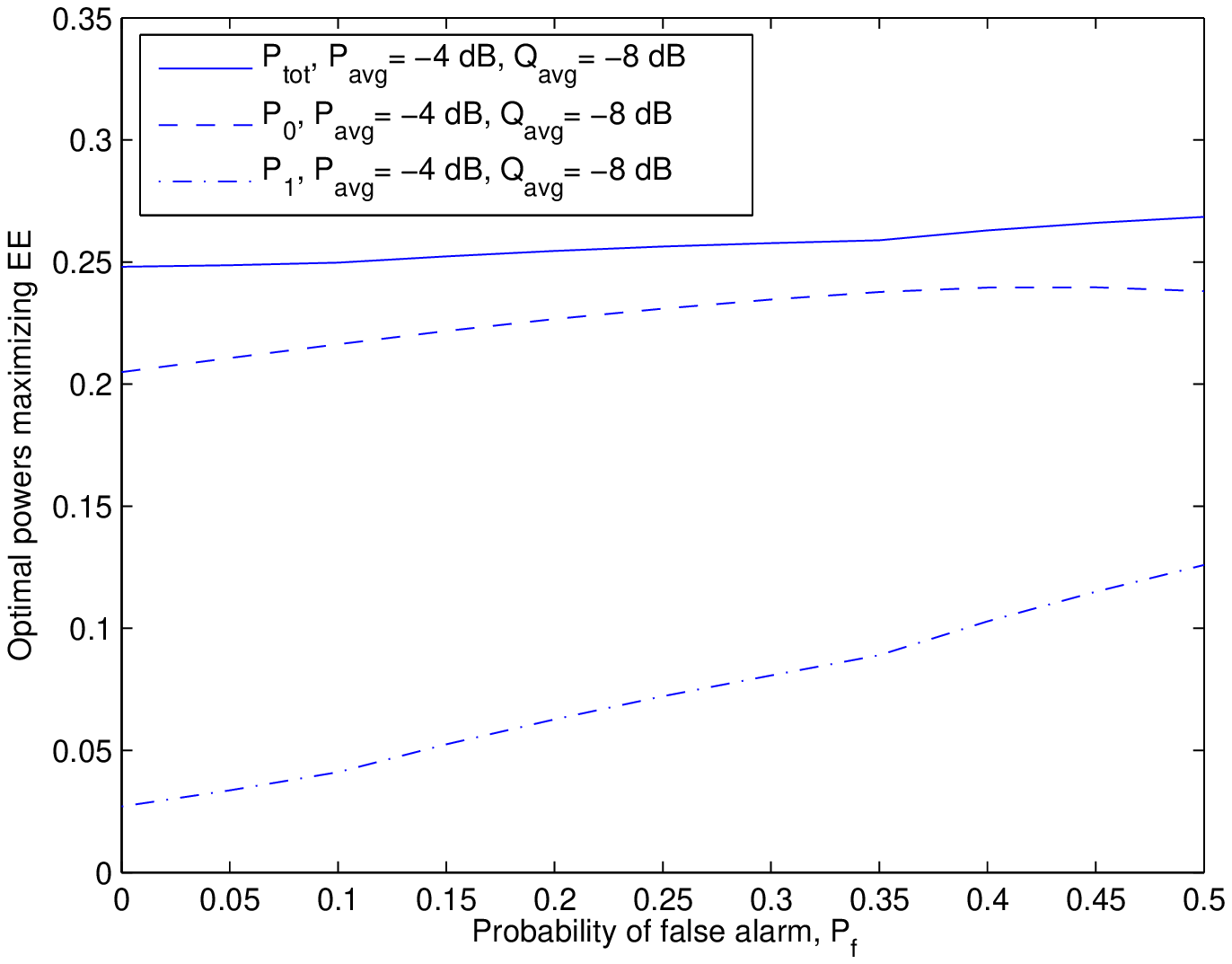}
\caption{$P_{\text{tot}}$, $P_0$ and $P_1$ vs. $P_{\f}$}
\end{subfigure}
\caption{\small{(a) Maximum achievable EE, $\eta_{\rm{EE}}(P_0,P_1)$ vs. probability of false alarm, $P_{\f}$; (b) optimal achievable rate maximizing EE, $R(P_0,P_1)$ vs. $P_{\f}$; (c) optimal total transmission power, $P_{\text{tot}}$ and $P_0$, $P_1$ vs. $P_{\f}$.}}\label{fig:EE_Pf}
\end{figure*}

In Fig. \ref{fig:EE_Pd}, maximum achievable EE, optimal achievable rate, $R(P_0,P_1)$, and optimal powers, $P_{\text{tot}}$, $P_0$ and $P_1$, are plotted as a function of the detection probability, $P_{\nid}$. We consider different peak and average transmit power constraints, e.g., $P_{pk,0}=P_{\pk,1}=P_{\avg} =-4$ dB and $P_{pk,0}=P_{\pk,1}=P_{\avg} = -10$ dB. Average interference constraint, $Q_{\avg}$ is set to $-8$ dB. It is assumed that probability of false alarm is $P_{\f}= 0.1$. We only display $R(P_0,P_1)$ and optimal powers, $P_{\text{tot}}$, $P_0$ and $P_1$, under the best performance, i.e., when $P_{\avg} =-4$ dB since the same trends are observed for other transmit power constraints. As $P_{\nid}$ increases, secondary users have more reliable sensing performance. Hence, secondary users experience miss detection events less frequently, which results in increased achievable rates. The transmission power $P_0$ under idle sensing decision increases with increasing $P_{\nid}$ while transmission power $P_1$ under busy sensing decision decreases with increasing $P_{\nid}$. Since the achievable rate increases and the total transmission power decreases, maximum achievable EE increases as sensing performance improves.

In Fig. \ref{fig:EE_Pf}, we display the maximum achievable EE, optimal achievable rate, $R(P_0,P_1)$, and optimal powers, $P_{\text{tot}}$, $P_0$ and $P_1$, as a function of the false alarm probability, $P_{\f}$. We again assume $P_{pk,0}=P_{\pk,1}=P_{\avg} = -4$ dB and $P_{pk,0}=P_{\pk,1}=P_{\avg} = -10$ dB. Since the optimal achievable rate and optimal powers, which maximize EE, show similar trends as a function of $P_{\f}$ subject to different peak and average transmit power constraints, we only consider the best performance achieved with $P_{\avg} =-4$ in Fig. \ref{fig:EE_Pf}(b) and Fig. \ref{fig:EE_Pf}(c). Average interference constraint is $Q_{\avg}=-8$ dB and the probability of detection, $P_{\nid}$, is set to $0.8$. As $P_{\f}$ increases, channel sensing performance deteriorates. Secondary users detect the idle channels as busy more frequently. Since the available channel is not utilized efficiently, secondary users have smaller achievable rates. Also, the total transmission power maximizing the EE increases with increasing $P_{\f}$, which leads to lower achievable EE.

\section{Conclusion}\label{sec:conc}
In this paper, we consider energy-efficient power adaptation for cognitive radio systems subject to peak/average transmit power constraints and average interference power constraints in the presence of sensing errors. EE maximization problem is transformed into an equivalent parameterized concave form and the optimal power adaptation schemes are derived. It is shown that power adaptation schemes depend on the sensing performance through detection and false alarm probabilities. Dinkelbach's method-based algorithm is proposed to iteratively solve the power allocation that maximizes the achievable EE. Numerically, we have several observations. For instance, it is shown that maximum achievable EE increases with increasing $P_{\nid}$ and decreases with increasing $P_{\f}$. Moreover, under the same average interference constraint, secondary users operating subject to peak transmit constraints have smaller achievable EE than that attained under average transmit power constraints.


\begin{thebibliography}{99}

\bibitem{fcc} Federal Communications Commission Spectrum Policy Task Force, ``FCC Report of the Spectrum Efficiency Working Group," Nov. 2002.

\bibitem{mitola} J. Mitola, \Rmnum{3}, ``Cognitive radio: An integrated agent architecture for software defined radio," Ph.D. dissertation, Royal Inst. Technol. (KTH), Stockholm, Sweden, 2000.

\bibitem{haykin} S. Haykin, ``Cognitive radio: Brain-empowered wireless communications," \emph{IEEE J. Sel. Areas Commun.}, vol. 23, no. 2, pp. 201-220, Feb. 2005.

\bibitem{gur} G. Gur and F. Alagoz, ``Green wireless communications via cognitive
dimension: an overview," \emph{IEEE Network}, vol. 25, pp. 50-56, Mar. 2011.

\bibitem{pei} Y. Pei, Y. -C. Liang, K. C. Teh, and K. H. Li, ``Energy-efficient design
of sequential channel sensing in cognitive radio networks: optimal sensing strategy, power allocation, and sensing order," \emph{IEEE J. Sel. Areas
Commun.}, vol. 29, no. 8, pp. 1648-1659, Sep. 2011.

\bibitem{shi} Z. Shi, K. C. Teh, and K. H. Li, ``Energy-efficient joint design of sensing and transmission durations for
protection of primary user in cognitive radio systems," \emph{IEEE Commun. Letters}, vol. 17, no. 3, pp. 565-568, Mar. 2013.

\bibitem{wang} S. Wang, M. Ge, and W. Zhao, ``Energy-efficient resource allocation for
OFDM-based cognitive radio networks," \emph{IEEE Trans. Commun.}, vol. 61, no. 8, pp. 3181-3191, Aug. 2013.


\bibitem{mao} J. Mao, G. Xie, J. Gao, and Y. Liu, ``Energy efficiency optimization for
cognitive radio MIMO broadcast channels," \emph{IEEE Commun. Letters}, vol. 17, no. 2, pp. 337-340, Feb. 2013.

\bibitem{ghasemi} A. Ghasemi and E. S. Sousa, ``Spectrum sensing in cognitive radio networks: requirements, challenges and design trade-offs,"
\emph{IEEE Comm. Mag.}, vol. 46, no, 4, pp. 32-39, Apr. 2008.

\bibitem{axell} E. Axell, G. Leus, E. G. Larsson, and H. V. Poor, ``Spectrum sensing for cognitive radio: State-of-the-art and recent advances," \emph{IEEE Signal Process. Mag.}, vol. 29, no. 3, pp. 101-116, May 2012.

\bibitem{ozcan} G. Ozcan and M. C. Gursoy, ``Achievable rate regions of cognitive multiple access channel with sensing errors," \emph{Proc. of the IEEE ISIT}, pp. 1660-1664, July 2013.

\bibitem{dinkelbach} W. Dinkelbach, ``On nonlinear fractional programming," \emph{Management Science}, vol. 13, pp. 492-498, Mar. 1967.

\bibitem{boyd} S. Boyd, L. Xiao, and A. Mutapcic, \emph{Subgradient methods}, Lecture Notes of EE392o, Standford University, Autumn Quarter 2003-2004.


\bibitem{schaible} S. Schaible, ``Fractional programming. \Rmnum{2}, On Dinkelbach's algorithm," \emph{Management Science}, vol. 22, no. 8, pp. 868-873, 1976.

\bibitem{kang} X. Kang, Y.- C. Liang, A. Nallanathan, H. K. Garg, and R. Zhang, ``Optimal power allocation for fading channels in cognitive radio networks: Ergodic capacity and outage capacity," \emph{IEEE Trans. Wireless Commun.}, vol. 8, no. 2, pp. 940-950, Feb. 2009.

\end{thebibliography}
\end{document}